\documentclass[11pt,twoside]{article}

\usepackage{asp2006}
\usepackage{epsf}
\usepackage{psfig}
\usepackage{lscape}

\begin{document}
\title{Non-equilibrium chemistry and dust formation in AGB stars as probed by SiO line emission}   
\author{Fredrik L. Sch{\"o}ier,\altaffilmark{1}
              Hans Olofsson,\altaffilmark{1,2}
              Tony Wong,\altaffilmark{3}
              David Fong,\altaffilmark{4}
              Michael Lindqvist,\altaffilmark{2} and
              Lorant O. Sjouwerman\altaffilmark{5}}
\altaffiltext{1}{Stockholm Observatory, AlbaNova University Center, SE-10691, Sweden}
\altaffiltext{2}{Onsala Space Observatory, SE-43992, Sweden}
\altaffiltext{3}{CSIRO Australia Telescope National Facility, PO Box 76, Epping NSW 1710, Australia}
\altaffiltext{4}{Harvard-Smithsonian Center for Astrophysics, 60 Garden Street, Cambridge, MA 02138, USA}
\altaffiltext{5}{National Radio Astronomy Observatory, PO Box O, Socorro, NM87801}

\begin{abstract} 
We have performed high spatial resolution observations of SiO line emission for a sample of 11 AGB stars using the ATCA, VLA and SMA interferometers. Detailed radiative transfer modelling suggests that there are steep chemical gradients of SiO in their circumstellar envelopes. The emerging picture is one where the radial SiO abundance distribution starts at an initial high abundance, in the case of M-stars consistent with LTE chemistry, that drastically decreases at a radius of $\sim$\,1\,$\times$\,10$^{15}$\,cm. This is consistent with a scenario where SiO freezes out onto dust grains. The region of the wind with low abundance is much more extended, typically $\sim$\,1\,$\times$\,10$^{16}$\,cm, and limited by photodissociation. The surpisingly high SiO abundances found in carbon stars requires non-equilibrium chemical processes. 

\end{abstract}


\section{Introduction}

Molecules can easily form in large abundance in the cool atmospheres of AGB stars and initiate a relatively complex chemistry that is further enhanced by photodissociation in the CSE. However,  most of the abundance estimates are based on rather simple methods and are typically order of magnitude estimates.
The first two more detailed studies of circumstellar abundances in larger samples of sources have been performed by \citet{Delgado03b} and \citet{Schoeier06a} for SiO in 45 M-type (C/O\,$<$\,1 in the photosphere) AGB stars and 19 carbon stars (C/O\,$>$\,1 in the photosphere), respectively. Average SiO fractional abundances were obtained from a detailed (non-local and NLTE) radiative transfer analysis of multi-transition single-dish observations. Interestingly, for the M-type AGB stars the
derived abundances are generally much lower than expected from photospheric equilibrium chemistry. For the carbon stars, on the other hand, the derived abundances are on the average two orders of magnitude higher than predicted by photospheric equilibrium chemistry.
Moreover,  there is a clear trend that the SiO fractional abundance decreases as the mass-loss rate of the star increases, as would be the case if SiO is accreted onto dust grains.

\begin{figure}[!ht]
\plotone{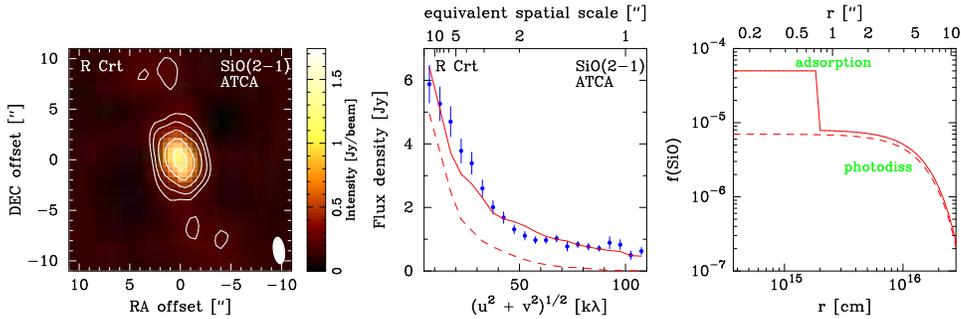}
\caption{Observed circumstellar SiO ($v$\,=\,0, $J$\,=\,2\,$\rightarrow$\,1) line emission and model results obtained for the M-type AGB star  R~Crt (Sch{\"o}ier et al., in prep.). The dashed line is the best fit single-component model determined from the single-dish data. The solid line is a two-component model (also consistent with the single-dish data) including a compact, pre-condensation, region with a high SiO abundance.}
\end{figure}

\section{Evidence of non-LTE chemistry and adsorption onto dust grains}

Recently, high-spatial-resolution interferometric SiO line observations have been published of the two M-type  AGB stars R~Dor and L$^2$~Pup by \citet{Schoeier04} and the carbon star IRC\,+10216 by \citet{Schoeier06b}. In addition, new SiO data have been obtained for IRAS 15194--5115, R Cas, IK Tau, R~Crt (see Fig.~1), IRC--10529, IRC+10365,  GX Mon, and W Hya using ATCA and VLA (Sch{\"o}ier et al., in prep.).
A detailed excitation analysis reveal the presence of an inner compact component of high fractional abundance (see Fig.~1; right panel), consistent with predictions from stellar atmosphere chemistry in the case of the M-type objects but several orders of magnitude larger than expected for the carbon stars, indicating the importance of non-LTE chemical processes as suggested by recent chemical models \citep{Cherchneff06}. In addition, an extended  low-abundance  component, as expected if SiO is effectively depleted onto grains in the inner wind, was required in order to fit the observations in all cases. 

Consequently, there are strong indications that circumstellar SiO line
emission carries information on the properties of the region where the
mass loss of AGB stars is initiated. However,  our knowledge of the relative importance of freeze-out onto dust grains, photodissociation, and circumstellar chemistry is still rudimentary. Higher spatial resolution observations and additional progress on chemical models, including grains, are needed for further progress.

\acknowledgements 
F.L.S. and H.O. acknowledge financial support from the Swedish research council.


\end{document}